\documentclass[12pt]{emulateapj}
\usepackage{apjfonts}

\newcommand{\pfrac}[2]{\left(\frac{#1}{#2}\right)}
\def\eps{\epsilon}
\def\ob{{\rm ob}}
\def\nukn{\nu^{\rm KN}}

\slugcomment{ }

\shorttitle{Preshock magnetic field of Fermi-LAT GRBs}
\shortauthors{Li \& Zhao}

\begin{document}

\title{The upstream magnetic field of collisionless GRB shocks: constraint by Fermi-LAT observations}
\author{Zhuo Li\altaffilmark{1,2} and Xiao-Hong Zhao\altaffilmark{3,4}}
 \altaffiltext{1}{Department of Astronomy, Peking University, Beijing 100871,
 China; E-mail:
zhuo.li@pku.edu.cn}
 \altaffiltext{2}{Kavli Institute for Astronomy and Astrophysics, Peking University, Beijing 100871,
 China}
 \altaffiltext{3}{National Astronomical Observatories/Yunnan
Observatory, Chinese Academy of Sciences, P.O. Box 110, 650011
Kunming, China}
 \altaffiltext{4}{Key Laboratory for the Structure and Evolution of Celestial Bodies,
Chinese Academy of Sciences, P.O. Box 110, 650011 Kunming, China }

\begin{abstract}
Long-lived $>100$~MeV emission has been a common feature of most
Fermi-LAT detected gamma-ray bursts (GRBs), e.g., detected up to
$\sim10^3$s in long GRBs 080916C and 090902B and  $\sim10^2$s in
short GRB 090510. This emission is consistent with being produced
by synchrotron emission of electrons accelerated to high energy by
the relativistic collisionless shock propagating into the weakly
magnetized medium. Here we show that this high-energy afterglow
emission constrains the preshock magnetic field to satisfy
$10^{0}n_0^{9/8}{\rm mG}< B_u<10^2n_0^{3/8}$mG, where $n_0$ is the
preshock density in unit of  $1~\rm cm^{-3}$, more stringent than
the previous constraint by X-ray afterglow observations on day
scale. This suggests that the preshock magnetic field is strongly
amplified, most likely by the streaming of high energy shock
accelerated particles.
\end{abstract}


\section{Introduction}
Diffusive (Fermi) shock acceleration is believed to play an
important role in gamma-ray burst (GRB) afterglow model, where a
shock propagating into the medium accelerates electrons to high
energy and then synchrotron/inverse-Compton emission arises.
Although the phenomenological afterglow model works generally well
with observational data, the shock physics is not understood from
first principle. One of the main issues is how the magnetic field
is amplified \citep[see review by][]{W06}. Many groups have tried
to address this issue using numerical plasma simulations
\citep[see,e.g.,][and references there in]{kkw09}, but because the
calculations are extremely demanding numerically, the picture of
field growth is still not clear. Thus using observations to
constrain the shock physics parameters will be helpful in this
issue. Afterglow observations were used to constrain the
downstream magnetic field and the accelerated electron energy
distribution, at both high \citep{W97a,fw01} and low
\citep{W97b,ew05} energy.  \citet[][hereafter LW06]{lw06} had also
used X-ray afterglow observations on day scale to give a lower
limit to the preshock magnetic field amplitude, which suggests
either the shock propagates into a magnetized wind of the GRB
progenitor or strong preshock magnetic field amplification occurs,
most likely due to the streaming of the accelerated high-energy
particles. LW06 further suggested that observations in higher
energy range and/or in later time may post more stringent
constraint on the preshock magnetic field.

Recently the Large Area Telescope (LAT) on board the Fermi
satellite revealed some new features in the GRB high-energy
emission. The onset of $>100$ MeV emission is delayed compared to
MeV emission, and lasts much longer, as a decaying power-law,
after the MeV emission already ends. For examples, in the bright
GRBs, the extended $>100$~MeV emission was detected up to
$\sim10^3$s in long GRBs 080916C and 090902B and $\sim10^2$s in
short GRB 090510 \citep{fermi09a,fermi09b,fermi09c}, until the
flux decreases to below the LAT sensitivity. \cite{kb09a,kb09b}
proposed that the whole high-energy burst, including the prompt
phase, is produced by synchrotron emission in the external forward
shock. Other authors
\citep{Corsi09,gao09,ghirlanda10,ghisellini10,de10,Pandey10} also
found that the light curve slope, the spectral index and the flux
level of the extended high-energy emission are consistent with the
synchrotron afterglow model. Even though it may not be true that
the whole burst is dominated by forward shock emission (see
discussion in \S\ref{sec:discussion}), for the $10^3$s-scale
emission the forward shock emission is still most favored over
other possible models \citep[see a discussion in the introduction
of][]{wang10}.

In this paper we show that if the extended $>100$~MeV emission is
produced by the synchrotron emission from afterglow shock
accelerated electrons, the $10^3$s-scale $>100$~MeV emission gives
much more stringent constraint on the preshock magnetic field,
compared to that by X-ray afterglow observations. The paper is
organized as follows: in \S2 we list the adopted preassumptions in
our analysis; in \S3 we derive the maximum synchrotron photon
energy limited by the radiative (IC and synchrotron/jitter)
cooling of the accelerated electrons; the KN correction to the IC
cooling rate is presented in \S4; then \S5 gives the constraint to
the upstream magnetic field; and the discussion on the result is
given in \S6. Unless stated otherwise, we use the common notation
$Q_x=Q/10^x$ and c.g.s. unit.

\section{Preassumption}\label{sec:assume}
We adopt some preassumptions in our following analysis.

(1) The $>100$~MeV emission at late time, $\ga10^2-10^3$s, is
assumed to be generated by the synchrotron emission in post-shock
(downstream) region by electrons accelerated in the afterglow
shock. Compared to other models, external-shock synchrotron model
provides the most natural explanation for the properties of the
extended emission, supporting this preassumption.

(2) The synchrotron photons are assumed to be IC scattered at most
once, i.e., neglecting the second-order IC scattering. This is
generally available since the electrons in early afterglow are
very energetic, their further interactions with up-scattered
synchrotron photons take place in deep Klein-Nishina (KN) regime.

The other minor preassumptions about the afterglow parameters are
below.

(3) The postshock (downstream) electrons' energy equipartition
parameter is $\eps_e\ga0.1$. This value is inferred from afterglow
modelling, and is consistent with the clustering of explosion
energy \citep{frail01} and X-ray afterglow luminosity
\citep{fw01}.

(4) The postshock magnetic field energy equipartition parameter is
in the range of $10^{-5}\la\eps_B\la10^{-2}$. In the afterglow
modelling, values of $10^{-3}\la\eps_B\la10^{-2}$ are usually
obtained \citep[e.g.,][]{pk01,harrison01}. Moreover, the X-ray
afterglow observations indicate $B_u>0.2n_0^{5/8}$~mG, with $n$
the medium density (LW06). An lower limit to the downstream
magnetic field strength is the shock-compressed field, $B'_{d,\rm
comp}=4\Gamma B_u$. Here $\Gamma$ is the Lorentz factor of
postshock plasma and the prime denotes the quantities in the frame
of the postshock plasma. Thus the lower limit to $\eps_B$ is
$\eps_{B,\rm comp}=B_{d,\rm comp}'^2/32\pi
\Gamma^2nm_pc^2=B_u^2/2\pi nm_pc^2$. With the lower limit to $B_u$
from X-ray observations, one has
\begin{equation}
  \eps_B>\eps_{B,\rm comp}\ga10^{-5}n_0^{1/4}
\end{equation}
(We have also discussed the case of much lower $\eps_B$, which
does not change much the final constraints; see
\S\ref{sec:Bulimit}). With these assumed values of $\eps_e$ and
$\eps_B$, we actually also have $\eps_e>\eps_B$, which is
consistent with the IC component observed in some afterglows
\citep[e.g., GRB 000926;][]{harrison01}.

(5) The medium density is assumed to be $10^{-2}\la n\la10^2\rm
cm^{-3}$. The $n$ value is not sure but this adopted range is
consistent with afterglow modelling
\citep[e.g.,][]{pk01,harrison01} and with our knowledge of the
interstellar medium.

(6) The postshock injected electrons follow a energy distribution
of $dn_e/d\gamma_e\propto\gamma_e^{-p}$ with the power-law index
assumed to be $2<p\la2.5$. These values are usually implied by the
afterglow observations \citep[e.g.,][and references
therein]{W97a,fw01}, and are consistent with the theoretical value
$p\approx2.2$ derived for isotropic diffusion of accelerated
particles (in the test particle limit) in both numerical
calculations \citep[e.g.,][]{acht01} and analytical analysis
\citep{kw05}.

\section{Diffusive shock acceleration and maximum synchrotron photon energy}
In the diffusive (Fermi) shock acceleration mechanism, high-energy
charged particles are scattered by upstream and downstream
magnetic field back and forth, respectively, crossing the shock
front multiples times, and gradually gain energy in each crossing.
The particle acceleration depends on the upstream and downstream
magnetic fields. We refer readers to LW06 for detailed discussion
and the formula used below.

For a relativistic shock expanding into the medium, with postshock
plasma of Lorentz factor $\Gamma\gg1$ and preshock magnetic field
$B_u$, the acceleration time of electrons to energy
$\gamma_em_ec^2$ is derived to be
\begin{equation}\label{eq:acctime}
  t_a=g\frac{\gamma_e'm_ec}{eB_u},
\end{equation}
where the correction factor $g$ accounts for that the upstream
electrons are not deflected by an angle of $1/\Gamma$ just with a
fraction $1/\Gamma$ of the Larmor time, because the preshock field
structure may not be uniform \citep[e.g.,][]{lp03,lr06} and that
the electrons do not return the downstream region once they are
deflected by $1/\Gamma$, because the shock velocity is decreasing
\citep[e.g.,][]{pn10}. A conservative value is adopted as
$g\approx10$ (LW06). Hereafter, primed variables denote parameter
values measured in the downstream frame, while non-primed ones in
the upstream frame.

\subsection{IC cooling}
When an electron resides in the upstream region, it may suffer
from energy loss due to IC scattering the synchrotron photons from
the downstream region. LW06 derived the IC cooling time of an
upstream-residing electron with $\gamma_e$, measured in the
downstream region, is
\begin{equation}\label{eq:ictime}
  t_c'=\frac{3m_ec}{4\sigma_Tu_{\rm ph}'\gamma_e'}.
\end{equation}
Here $u_{\rm ph}'$ is the synchrotron radiation energy density
measured in the downstream frame. Because the electrons are
confined in a region with thickness much smaller than the size of
the shock $R$, the synchrotron radiation energy densities in the
upstream and downstream regions that we concern are similar, thus
$u_{\rm ph}'$ can be either of them.

The IC scattering may occur in KN regime, then $u_{\rm ph}'$
should be considered as the "effective" energy density of
scattered photons. Let us define the photon frequency
\begin{equation}
  \nukn_x\equiv\nukn(\gamma_x)=\frac{\Gamma m_ec^2}{h\gamma'_x}
\end{equation}
so that the IC scattering between electrons with
(downstream-frame) Lorentz factor $\gamma_x'$ and photons at
$\nu>\nukn_x$ enters KN regime. Except for the spectrum of the
seed photons are very hard (which is not the case here), an
electron of $\gamma_x$ mainly interacts with photons below
$\nukn_x$. So if KN effect is important, the photon energy density
in eq.(\ref{eq:ictime}) should be replaced by that below
$\nukn(\gamma_e)$, $u_{\rm ph}'\rightarrow u_{\rm ph}'(<\nukn)$.

An electron that has been accelerated to $\gamma_em_ec^2$ must
satisfy that its upstream IC cooling time is longer than the
acceleration time, $t_c'>t_a/\Gamma$, which gives an upper limit
to $\gamma_e'$,
\begin{equation}\label{eq:gammamaxic}
  \gamma'_e<\pfrac{3eB_u\Gamma}{4\sigma_Tgu_{\rm
  ph}'(<\nukn)}^{1/2}.
\end{equation}

For the following derivation, it is convenient to define the
"effective" downstream Compton parameter $Y_x$ for relevant
electrons of $\gamma_x$ as
\begin{equation}\label{eq:Yapproximation}
  Y_x=\frac{u_{\rm ph}'(<\nukn_x)}{u_B'},
\end{equation}
with $u_B'=B_d'^2/8\pi$ being the energy density of postshock
magnetic field, and further replace $u_{\rm ph}'$ in
eq.(\ref{eq:gammamaxic}) with $u_{\rm
ph}'(<\nukn)=Yu'_B$\footnote{Eq. (\ref{eq:Yapproximation}) is an
approximation, however, as shown in the Appendix, considering the
full KN effect (Eq. [\ref{eq:Y_KN}]) only gives a correction
within 20\% in the problem.}.

The characteristic frequency of synchrotron photons emitted by
electrons with $\gamma_e'$ downstream is
$\nu=0.3\Gamma\gamma_e'^2eB'_d/2\pi m_ec$, where it has been taken
into account that the synchrotron radiation peaks at 0.3 times the
gyration frequency of relevant electrons. Thus the upper limit in
eq. (\ref{eq:gammamaxic}) for $\gamma_e'$ implies also a maximum
observed energy of synchrotron photons,
\begin{equation}\label{eq:iclimitenergy}
  h\nu_\ob<7.1\frac{B_{u,-6}E_{54}^{1/8}}{\eps_{B,-2}^{1/2}g_1n_0^{5/8}t_3^{3/8}Y_\ob}\rm
  keV,
\end{equation}
where $Y_\ob$ is the effective Compton parameter of electrons with
$\gamma_\ob'$ that emitting synchrotron photons at $\nu_\ob$. Here
$E$ is the total (isotropic-equivalent) kinetic energy of the
shock, $n$ is the medium density, and the downstream field $B_d'$
is calculated from equipartition as $B_d'=(32\pi\eps_B
nm_p)^{1/2}\Gamma c$, where one still needs dynamical evolution of
the shock. In late time when the shock follows the Blandford-McKee
self-similar solution, the Lorentz factor $\Gamma$ drops with
radius $R$ as $\Gamma=(17E/16\pi nm_pc^2)^{1/2}R^{-3/2}$
\citep{bm76}. Taking the equal arrival time surface into account,
the relation between $\Gamma$ and observer time $t$ is
$t=R/4\Gamma^2c$ \citep{W97c}.

We will calculate later in \S\ref{sec:KN} the value of $Y_\ob$ as
function of a range of $\eps_B$ value, considering the KN
correction for IC cooling of electrons.

\subsection{Synchrotron/jitter cooling}
During the upstream residence time, the electrons will also suffer
energy loss due to synchrotron or jitter radiation when gyrating
or being deflected by the upstream field. The energy loss rate of
an electron of $\gamma_e=\Gamma\gamma'_e$ due to synchrotron and
jitter radiation in the upstream field $B_u$ is
$\dot{E}=(4/3)\sigma_Tc\gamma_e^2(B_u^2/8\pi)$, then the cooling
time due to this process (measured in the upstream frame) is
$t_B=\gamma_em_ec^2/\dot{E}$, i.e.,
\begin{equation}
  t_B=\frac{6\pi m_ec}{\sigma_T\gamma_e'\Gamma B_u^2}.
\end{equation}
The electron successfully accelerated to $\gamma_em_ec^2$, again,
requires its energy loss timescale larger than the acceleration
time (eq.[\ref{eq:acctime}]), $t_B>t_a$, which gives another limit
to $\gamma_e$,
\begin{equation}
  \gamma'_e<\pfrac{6\pi e}{\sigma_TgB_u\Gamma}^{1/2}.
\end{equation}
Therefore another limit to the observed energy of synchrotron
photons is
\begin{equation}\label{eq:synlimitenergy}
  h\nu_\ob<11\frac{E_{54}^{1/8}\eps_{B,-2}^{1/2}n_0^{3/8}}{B_{u,-6}g_1t_3^{3/8}}\rm
  TeV.
\end{equation}

\section{KN correction to IC cooling}\label{sec:KN}
We are interested in those electrons emitting synchrotron photons
in the LAT range. We should take $h\nu_\ob=100$~MeV and then
$\gamma_\ob'$ and $Y_\ob$ are the Lorentz factor and effective
Compton parameter of relevant electrons, respectively. Now we
derive the value of $Y_\ob$, considering the KN correction. For
this purpose we should derive the synchrotron spectral
distribution and then calculate the energy density of the
synchrotron radiation below $\nukn_\ob$. As $\eps_e\gg\eps_B$, IC
cooling may dominate synchrotron cooling. If the IC scattering
takes place in KN regime, the complexity arises since in this case
the energy distribution of the postshock electrons and the
synchrotron photons are coupled and affect each other.

The injected electron distribution downstream follows a power law
$\propto\gamma_e^{-p}$ with minimum Lorentz factor
$\gamma_m'=f_p\langle\gamma_e'\rangle=\eps_ef_p(m_p/m_e)\Gamma$.
Here $f_p$ is the ratio between $\gamma_m'$ and the average
Lorentz factor. As $p>2$ the electron energy is dominated by the
low-energy end electrons, thus $f_p<1$. If the electron
distribution at the low energy end is an abrupt cutoff at
$\gamma_m$, then $f_p=(p-2)/(p-1)$. A smoother turnover at
$\gamma_m'$ will lead to $f_p>(p-2)/(p-1)$. The synchrotron
radiation by electrons with $\gamma_m'$ peaks at frequency
$\nu_m=0.3\Gamma\gamma_m'^2eB'_d/2\pi m_ec$, i.e.,
\begin{equation}\label{eq:num}
  \nu_m=1.3\times10^{16}E_{54}^{1/2}\eps_{B,-2}^{1/2}\eps_{e,-1}^2f_p^2t_3^{-3/2}\rm
  Hz.
\end{equation}
We focus mainly on the point when $t\approx10^3$s. Electrons with
$\gamma_m'$ mainly interact with synchrotron photons below
frequency $\nukn_m$, where
\begin{equation}\label{eq:omtonum}
  \frac{\nukn_m}{\nu_m}=54E_{54}^{-1/2}\eps_{B,-2}^{-1/2}\eps_{e,-1}^{-3}f_p^{-3}t_3^{3/2}.
\end{equation}

At high enough electron energy, the radiative (synchrotron and IC)
loss time is shorter than the adiabatic cooling time, i.e, the
energy-loss time due to the expansion of the postshock plasma. The
adiabatic loss time, $t_{\rm ad}'=6R/13c\Gamma$
\citep{Gruzinov99}, is longer than the radiative cooling time for
electrons with Lorentz factors exceeding
$\gamma_c'=3m_ec/4\sigma_Tu_B'(1+Y_c)t_{\rm ad}'$. Here $Y_c$ is
the Compton $Y$-parameter for electrons with
$\gamma_e'=\gamma_c'$. The characteristic synchrotron frequency of
photons emitted by electrons with $\gamma_e'=\gamma_c'$,
$\nu_c=0.3\Gamma\gamma_c'^2eB'_d/2\pi m_ec$, is
\begin{equation}\label{eq:nuc}
  \nu_c=2.2\times10^{14}E_{54}^{-1/2}\eps_{B,-2}^{-3/2}n_0^{-1}t_3^{-1/2}(1+Y_c)^{-2}\rm
  Hz.
\end{equation}
Electrons with $\gamma_e'=\gamma_c'$ mainly interact with
synchrotron photons below a frequency
\begin{equation}\label{eq:oc}
  \nukn_c=5.1\times10^{18}E_{54}^{1/2}\eps_{B,-2}n_0^{1/2}t_3^{-1/2}(1+Y_c)\rm
  Hz.
\end{equation}
So we also have
\begin{equation}\label{eq:octonuc}
  \frac{\nukn_c}{\nu_c}=2.3\times10^4E_{54}\eps_{B,-2}^{5/2}n_0^{3/2}(1+Y_c)^3.
\end{equation}

For those electrons with $\gamma_e'=\gamma_\ob'$ that emit
synchrotron photons at $h\nu_\ob=100$~MeV
($\nu_\ob=0.3\Gamma\gamma_\ob'^2eB'_d/2\pi m_ec$), the
corresponding KN frequency is
\begin{equation}\label{eq:oob}
  \nukn_\ob=4.9\times10^{14}E_{54}^{1/4}\eps_{B,-2}^{1/4}t_3^{-3/4}\rm
  Hz,
\end{equation}
thus we further have
\begin{equation}\label{eq:oobtonum}
  \frac{\nukn_\ob}{\nu_m}=3.9\times10^{-2}E_{54}^{-1/4}\eps_{B,-2}^{-1/4}\eps_{e,-1}^{-2}f_p^{-2}t_3^{3/4}.
\end{equation}

Synchrotron self-absorption may be important at low frequencies.
Consider the extreme case where all the postshock electrons emit
at the absorption frequency $\nu_a$ with the electron Lorentz
factor being $\gamma_a'=(2\pi m_ec\nu_a/0.3\Gamma eB_d')^{1/2}$.
The absorption coefficient is then $\alpha_\nu\approx4\Gamma
ne^3B_d'/2\gamma_a'(m_ec\nu_a)^2$. Through $\alpha_\nu
R/\Gamma\approx1$ we obtain
\begin{equation}
  \nu_a\approx9.5\times10^{11}E_{54}^{1/5}\eps_{B,-2}^{3/10}n_2^{1/2}t_3^{-1/5}\rm
  Hz.
\end{equation}
Note that this should be taken as the upper limit to $\nu_a$
because what we assume is the "extreme case" where the
self-absorption is strongest, and $\eps_{B}$ and $n$ values have
been plugged with the maximum ones. Compared with the KN frequency
we have $\nukn_\ob\gg\nu_a$, then reach the conclusion that the
synchrotron self-absorption is negligible.

There are uncertainties in the afterglow model parameters,
especially for the poorly constrained $\eps_B$, which is usually
coupled with $n$ in the afterglow modelling. We should scan all
the possible parameter space. However, when IC cooling is
important and KN correction is important to IC cooling, the
electron distribution (after cooling modified) and synchrotron
spectrum become quite complicated compared to the cases when
synchrotron cooling dominates or IC scattering takes place in
Thomson limit \citep[e.g.,][]{nakar09,wz09,wang10}. It is better
to pin down some relations between the characteristic frequencies
in the spectrum, and cancel the other cases in the parameter
space.

As we are going to consider the $>100$~MeV emission of $10^3$s
scale in GRBs 080916C and 090902B (with isotropic equivalent
gamma-ray energy $E_\gamma$ of order $10^{54}$erg) and of $10^2$s
scale in GRB 090510 ($E_\gamma\sim10^{53}$erg), we will consider
only two cases with $(E,t)=(10^{54}{\rm erg},10^3{\rm s})$ and
$(10^{53}{\rm erg},10^2{\rm s})$. In these two cases, with the
presumed parameter ranges mentioned in \S\ref{sec:assume}, we
examine the above calculations and find the followings are always
satisfied,
\begin{equation}\label{eq:cond1}
  \nukn_\ob<\nu_m<\nukn_m
\end{equation}
and
\begin{equation}\label{eq:cond2}
  \nukn_\ob<\nukn_c~~(\nu_\ob>\nu_c).
\end{equation}

Depending on the relation between $\nu_m$ and $\nu_c$, there are
two regimes for the bulk postshock injected electrons: "slow
cooling" regime with $\nu_m<\nu_c$ and "fast cooling" regime with
$\nu_c<\nu_m$. Consider the critical case when $\nu_m=\nu_c$, then
the synchrotron emission is peaking at $\nu_m(=\nu_c)$. Given
$\nukn_m>\nu_m$, the IC scattering does not suffer KN suppression,
and if only single IC scattering is considered, then
$Y_m=Y_c=(\eps_e/\eps_B)^{1/2}$ \citep{se01}, which substituted
into eq.(\ref{eq:nuc}) and $\nu_m=\nu_c$ leads to a critical
$\eps_B$ value
\begin{equation}
  \eps_{B,\rm
  cr}=1.7\times10^{-5}\frac{t_3}{E_{54}\eps_{e,-1}^3f_p^2n_0}.
\end{equation}
For $\eps_{B}<\eps_{B,\rm cr}$ electrons are slow cooling, and
vice verse. As the wide parameter space may allow both regimes to
happen, in what follows we derive $Y_\ob$ in these two regimes
separately.

\subsection{Slow cooling regime}
In this case the $\nu f_\nu$ spectrum of the synchrotron radiation
peaks at $\nu_c$, i.e., the total energy density in synchrotron
photons is $u_{\rm ph,syn}'\approx u_{\rm ph}'(<\nu_c)$. Let us
first derive $Y_c$.

As only the electrons with $\gamma_e'>\gamma_c'$ efficiently cool,
let us denote $\eta\equiv(\nu_m/\nu_c)^{(p-2)/2}<1$ the fraction
of postshock injected electron energy that is rapidly radiated
\citep{se01}. If $p\approx2$ then the value of $\eta$ is usually
order of unity. A lower limit can be obtained by taking $Y_c=0$,
then eqs. (\ref{eq:num}) and (\ref{eq:nuc}) implies
$\eta>0.4(E_{54}\eps_{B,-5}^2\eps_{e,-1}^2f_p^2n_0/t_3)^{0.1}$ for
$p=2.2$.

As the synchrotron spectrum follows $\nu f_\nu\propto \nu^{4/3}$
at $\nu<\nu_m$, $\propto\nu^{(3-p)/2}$ at $\nu_m<\nu<\nu_c$, and
then decreases with $\nu$ above $\nu_c$. The spectral shape above
$\nu_c$ is affected by KN correction, but is not interested to
this work. Following \cite{wang10}, we can set up equations for
$Y_c$, depending on the relations between $\nukn_c$, $\nu_m$ and
$\nu_c$:
\begin{equation}\label{eq:slowcoolingYc}
  Y_c(1+Y_c)=\frac{\eta\eps_e}{\eps_B}\left\{
  \begin{array}{ll}
    \pfrac{\nu_m}{\nu_c}^{(3-p)/2}\pfrac{\nukn_c}{\nu_m}^{4/3}c_1
    & \nukn_c<\nu_m<\nu_c,\\
    \pfrac{\nukn_c}{\nu_c}^{(3-p)/2}c_2 & \nu_m<\nukn_c<\nu_c,\\
    c_3 &  \nu_c<\nukn_c,
  \end{array} \right.
\end{equation}
where, assuming $\nu f_\nu\propto\nu^{-(p-2)/2}$ for $\nu>\nu_c$,
the correction factors are, approximately,
$c_1\approx{3\over8}(3-p)(p-2)$ if $\nukn_c<\nu_m\ll\nu_c$,
$c_2\approx p-2$ if $\nu_m\ll\nukn_c<\nu_c$ and $c_3\approx1$ if
$\nu_c\ll\nukn_c$. Now substituting eqs. (\ref{eq:num}),
(\ref{eq:nuc}) and (\ref{eq:oc}) into eq.(\ref{eq:slowcoolingYc}),
$Y_c$ can be solved out.

Since in the parameter space that we concern, we have conditions
of eqs. (\ref{eq:cond1}) and (\ref{eq:cond2}), there are only
several interesting cases for $Y_\ob$:
\begin{equation}\label{eq:slowcoolingYob}
  Y_\ob= Y_c\left\{
  \begin{array}{ll}
    \pfrac{\nukn_\ob}{\nukn_c}^{4/3}c_4  & \nukn_\ob<\nukn_c<\nu_m<\nu_c,\\
    \pfrac{\nukn_\ob}{\nu_m}^{4/3}\pfrac{\nu_m}{\nukn_c}^{(3-p)/2}c_5  & \nukn_\ob<\nu_m<\nukn_c<\nu_c,\\
    \pfrac{\nukn_\ob}{\nu_m}^{4/3}\pfrac{\nu_m}{\nu_c}^{(3-p)/2}c_6  &
    \nukn_\ob<\nu_m<\nu_c<\nukn_c,
  \end{array} \right.
\end{equation}
where the correction factors are approximately $c_4\approx1$ if
$\nukn_\ob<\nukn_c<\nu_m\ll\nu_c$, $c_5\approx{3\over8}(3-p)$ if
$\nukn_\ob<\nu_m\ll\nukn_c<\nu_c$, and
$c_6\approx{3\over8}(3-p)(p-2)$ if
$\nukn_\ob<\nu_m\ll\nu_c\ll\nukn_c$.

For the case with the lowest $\eps_B$ value allowed,
$\eps_B=10^{-5}$, and with the other parameters being
$E_{54}=t_3=n_0=\eps_{e,-1}=1$, the condition for slow cooling
regime is marginally satisfied, $\eps_B\la\eps_{B,\rm cr}$. We
derive $Y_\ob$ to be, as shown in Appendix,
\begin{equation}\label{eq:Yob_eq:eB=1e-5}
  Y_\ob\approx0.99\frac{c_3c_6}{0.1}\frac{\eta E_{54}^{1/6}n_0^{1/2}
  t_3^{1/2}}{\eps_{B,-5}^{1/3}\eps_{e,-1}^{2/3}f_p^{5/3}},
\end{equation}
where $p=2$ is used when $p$ appears in indices. We neglect the
redshift effect so far. Considering this effect, i.e.,
$\nu_\ob\rightarrow\nu_\ob(1+z)$ and $t\rightarrow t/(1+z)$, the
r.h.s. of eq. (\ref{eq:Yob_eq:eB=1e-5}) should be multiplied by
$(1+z)^{-7/6}$.

\subsection{Fast cooling regime}
When $\eps_B>\eps_{B,\rm cr}$ is satisfied, we have $\nu_c<\nu_m$
($\gamma_c'<\gamma_m'$). This means all the postshock injected
electron energy is radiated rapidly, $\eta=1$.

Since in the parameter space we are interested we have
$\nukn_m>\nu_m$, the electrons with $\gamma_e'=\gamma_m'$ do not
suffer KN suppression in IC scattering the synchrotron photons,
and the synchrotron $\nu f_\nu$ spectrum is peaking at $\nu_m$.
Now that $\gamma_m>\gamma_c$, we have $\nukn_c>\nukn_m>\nu_m$,
i.e., electrons around $\gamma_c'$ even have less KN correction in
IC scattering synchrotron photons. In this case, the electron
distribution at $\gamma_c<\gamma_e<\gamma_m$ still follows the
result derived in Thomson limit,
$dn_e/d\gamma_e\propto\gamma_e^{-2}$ \citep{se01}. The relevant
synchrotron spectrum follows $\nu f_\nu\propto\nu^{4/3}$ at
$\nu<\nu_c$, $\propto\nu^{1/2}$ at $\nu_c<\nu<\nu_m$, and then
turnover at $\nu_m$, with $u'_{\rm ph,syn}\approx u'_{\rm
ph}(<\nu_m)$. Without KN effect, the Compton parameter $Y_m$ of
electrons with $\gamma_m'$ is still the same as the solution in
Thomson limit \citep{se01},
\begin{equation}\label{eq:fastcoolYm}
  Y_m= Y_c=\pfrac{\eps_e}{\eps_B}^{1/2}.
\end{equation}
Using the spectral form of synchrotron radiation and depending on
the relations of $\nukn_\ob$ with $\nu_c$ and $\nu_m$, the Compton
parameter of electrons with $\gamma_\ob'$ is given by
\begin{equation}\label{eq:fastcoolingYob}
  Y_\ob= Y_m\left\{\begin{array}{ll}
    \pfrac{\nukn_\ob}{\nu_c}^{4/3}\pfrac{\nu_c}{\nu_m}^{1/2}c_7  &
    \nukn_\ob<\nu_c<\nu_m,\\
    \pfrac{\nukn_\ob}{\nu_m}^{1/2}c_8  &  \nu_c<\nukn_\ob<\nu_m,
  \end{array}\right.
\end{equation}
where $c_7\approx{3\over8}(p-2)/(p-1)$ if
$\nukn_\ob<\nu_c\ll\nu_m$, and $c_8\approx(p-2)/(p-1)$ if
$\nu_c\ll\nukn_\ob<\nu_m$. The case of $\nu_m<\nukn_\ob$ is
neglected provided the condition of eq. (\ref{eq:cond1}). Thus
with helps of eqs. (\ref{eq:num}), (\ref{eq:nuc}), (\ref{eq:oob})
and (\ref{eq:fastcoolYm}), we now can calculate $Y_\ob$.

In the case of high $\eps_B$ value, $\eps_{B}=10^{-2}$, with the
other parameters being $E_{54}=t_3=n_0=\eps_{e,-1}=1$, we get, see
Appendix,
\begin{equation}
  Y_\ob\approx0.12\frac{c_8}{0.2}\frac{t_3^{3/8}}{E_{54}^{1/8}\eps_{B,-2}^{5/8}\eps_{e,-1}^{1/2}f_p}.
\end{equation}
Here the redshift effect will add a term $(1+z)^{-5/8}$ on the
r.h.s..

We have also calculated the $Y_\ob$ value in a while parameter
space of $\eps_B=10^{-5}-10^{-2}$ and $n=10^{-2}-10^2\rm cm^{-3}$
for both cases of $(E,t)=(10^{54}{\rm erg},10^3{\rm s})$ and
$(10^{53}{\rm erg},10^2{\rm s})$, with the results presented in
Table \ref{tab}.

\begin{table}
\caption{Compton $Y$-parameter of electrons emitting 100-MeV
synchrotron photons}
\begin{center}
\begin{tabular}{|ll|lll|ll|}
\hline &  & \multicolumn{3}{|c|}{$E_{54}=t_3=1$} &
\multicolumn{2}{|c|}{$E_{53}=t_2=1$} \\
 &  & \multicolumn{3}{|c|}{$n(\rm cm^{-3})$} &
\multicolumn{2}{|c|}{$n(\rm cm^{-3})$ } \\
&  &  $10^{-2}$  & 1 & $10^2$ & $10^{-2}$ & 1 \\\hline
& $10^{-2}$ & $0.02c_{7,-1}$ & $0.06c_{8,-1}$  &  $0.06c_{8,-1}$  &  $0.004c_{7,-1}$  &  $0.03c_{8,-1}$ \\
\raisebox{1.5ex}[0pt]{$\eps_B$} & $10^{-5}$ & $0.1c_{25,-1}$  & $1c_{36,-1}$  &  $10c_{36,-1}$  &  $0.02c_{14,-1}$  & $0.2c_{36,-1}$ \\
\hline
\end{tabular}
\end{center}
\par
\tablecomments{The other parameters are $\eps_e=0.1$, $f_p=1$ and
$p=2$ if $p$ appears in exponents. Here $c_{14}=c_1c_4$,
$c_{25}=c_2c_5$, $c_{36}=c_3c_6$, and $c_{x,-1}=c_x/10^{-1}$.}
\label{tab}
\end{table}

\section{Constraint on preshock magnetic field}\label{sec:Bulimit}
For LAT energy range, we should take $h\nu_\ob=$100 MeV in
eq.(\ref{eq:iclimitenergy}), which, then, gives a lower limit to
the upstream field,
\begin{equation}\label{eq:Buic}
  B_u>14\frac{\eps_{B,-2}^{1/2}g_1n_0^{5/8}t_3^{3/8}Y_\ob}{E_{54}^{1/8}}\pfrac{h\nu_\ob}{\rm 100~MeV}(1+z)^{5/8}\rm
  mG.
\end{equation}
Note redshift effect is added hereinafter, and $Y_\ob$ has been
derived in \S\ref{sec:KN}. Plugging the $Y_\ob$ values calculated
in the cases of $(E,t)=(10^{54}{\rm erg},10^3{\rm s})$ and
$(10^{53}{\rm erg},10^2{\rm s})$ into eq.(\ref{eq:Buic}), it is
easy to find that the constraint by the latter case is less
stringent, i.e., the value of lower limit to $B_u$ is smaller.
Moreover, in the former case, we have
\begin{eqnarray}\label{eq:Blower}
   B_u(\eps_B=10^{-5})>0.5\frac{g_1t_3^{3/8}}{E_{54}^{1/8}}\frac{Y_\ob}{1.2}n_0^{5/8}(1+z)^{5/8}{\rm
   mG},   \\
   B_u(\eps_B=10^{-2})>2\frac{g_1t_3^{3/8}}{E_{54}^{1/8}}\frac{Y_\ob}{0.12}n_0^{5/8}(1+z)^{5/8}{\rm
   mG}.
\end{eqnarray}
The result of more accurate calculation in $\eps_B=10^{-5}$ case,
$Y_\ob(z=0)=1.2$ (see Appendix), has been used. We see that the
relatively relaxed lower limit to $B_u$ appears when taking the
lowest allowed value, $\eps_B=10^{-5}$, but the limit is in the
order of mG, and insensitive to the $\eps_B$ value.

One may argue that the $\eps_B\ga10^{-5}$ assumption is motivated
by X-ray afterglow observations on day scale, and may not hold at
early time, $\sim10^3$s, that is addressed in this paper. Indeed,
if relaxing the $\eps_B>10^{-5}$ assumption, the lower limit to
$B_u$ does not change much, and only varies slowly with $\eps_B$
as $\sim\eps_{B}^{1/6}$.

Furthermore, with $h\nu_\ob=100$~MeV, eq.(\ref{eq:synlimitenergy})
gives an upper limit to the upstream field,
\begin{equation}\label{eq:Busyn}
  B_u<110\frac{E_{54}^{1/8}\eps_{B,-2}^{1/2}n_0^{3/8}}{g_1t_3^{3/8}}\pfrac{h\nu_\ob}{\rm 100~MeV}^{-1}(1+z)^{-5/8}\rm
  mG.
\end{equation}
It can be seen that taking $(E,t)=(10^{54}{\rm erg},10^3{\rm s})$
gives more stringent constraint than $(E,t)=(10^{53}{\rm
erg},10^2{\rm s})$. As we expect $\eps_B\la10^{-2}$, then we have
the relatively relaxed constraint
\begin{equation}\label{eq:Bupper}
  B_u(\eps_B=10^{-2})<10^2\frac{E_{54}^{1/8}}{g_1t_3^{3/8}}n_0^{3/8}(1+z)^{-5/8}\rm
  mG.
\end{equation}

It is interesting to note that the constraints by IC and
synchrotron/jitter cooling together limit the upstream field in a
closed range that spans about two orders of magnitude.

\section{Discussion}\label{sec:discussion}
We have shown that $10^3$s-scale, $>100$~MeV GRB emission recently
revealed by Fermi-LAT provides stringent constraints on the
upstream magnetic field. A lower limit to the magnetic field is
obtained by requiring the acceleration time of electrons producing
$>100$~MeV synchrotron photons to be shorter than their
energy-loss time due to IC scattering the afterglow photons. By
scanning the possible afterglow parameter space, the lower limit
for the magnetic field is given in eq. (\ref{eq:Blower}).
Interestingly, an upper limit to the magnetic field is also
obtained by requiring the acceleration time to be shorter than the
energy-loss time due to synchrotron/jitter radiation upstream.
Given a maximum equipartition value of magnetic field downstream,
the upper limit for the field is given in eq. (\ref{eq:Bupper}).
Combining both lower and upper limits, the upstream magnetic field
is limited in a closed range with two orders of magnitude
uncertainty, $10^0n_0^{9/8}{\rm mG}\la B_u\la10^2n_0^{3/8}$mG.

There should be another lower limit to $B_u$ by requiring that the
acceleration time of 100-MeV emitting electrons is shorter than
the dynamical time of the afterglow shock. As LW06, we neglect
this "age limit" because the constraint is much less stringent
than the one by IC cooling. Even assuming the $\sim10$ GeV photons
from the Fermi-LAT GRBs is produced by external shock synchrotron
radiation, the constraint to $B_u$ by this argument, $B_u>0.1$~mG
\citep{pn10}, is still much less stringent than the cooling limit
here.

The lower limit to upstream field by Fermi-LAT observations is
larger than the previous constraint using X-ray afterglow
observations (LW06) by orders of magnitude. It can be seen that
this high amplitude field is not likely to be provided by a
magnetized wind from the GRB progenitor (see discussion in LW06),
therefore the only reasonable origin of this field is due to
magnetic field amplification upstream, most likely by the
streaming of high energy shock accelerated particles
\citep[e.g.,][]{Couch08,Medvedev09,Pelletier09,Lemoine10,Niemiec10}.
However, compared to the common few~$\mu$G-scale interstellar
medium field, this means an amplification of field amplitude by at
least 3 orders of magnitude, $\delta B/B>10^3$, or amplification
of field energy density by $>6$ orders of magnitude, $\delta
u_B/u_B>10^6$. Such high contrast amplification suggests that the
amplification may have nothing to do with the prior interstellar
medium field. This supports a co-evolution picture for magnetic
field and accelerated particles in the shock.

The lower limit to the upstream field may also imply a lower limit
to the downstream one, i.e., the shock-compression field. The
downstream field limit is, then, $\eps_{B,\min}=\eps_{B,\rm
comp}=B_u^2/2\pi nm_pc^2$. The upstream limit assuming
$\eps_{B,\rm ass}=10^{-5}$ is $B_u>0.5n_0^{9/8}$mG, which however
implies a larger downstream limit
$\eps_{B,\min}=4\times10^{-5}n_0^{5/4}>\eps_{B,\rm ass}$, thus
there is no self-consistence in the case of $\eps_B=10^{-5}$. The
present lower limit to the upstream field is mG scale for a wide
range of $\eps_B$, which implies shock compressed downstream field
of $\eps_{B,\rm comp}\approx10^{-4}$, therefore to be
self-consistent it is required that $\eps_B>10^{-4}$. This is
consistent with the results from afterglow modelling
\citep[e.g.,][]{pk01,harrison01}. Besides, the upper limit to
$B_u$ is close to the maximum value by equipartition argument,
$B_u^2/8\pi\approx nm_pc^2$. So the upstream field might be
amplified to be close to equipartition, although still dominated
by preshock matter in energy.

An important assumption in our analysis is that the long-lived,
$>10^3$s, LAT energy range emission is produced by synchrotron
radiation from electrons accelerated by afterglow shocks. We
noticed very recently there is a debate on the external shock
interpretation for the LAT detected GeV emission
\citep{pn10,bk10}. We note here that the debate is not relevant to
our assumption. The argument against external shock model
\citep{pn10} concerns the very high-energy, $>10$~GeV, photon
arrives at $\sim10^2$s, as observed in GRB 090902B
\citep{fermi09b}. However we consider the lower-energy,
$>100$~MeV, emission of longer timescale, $\ga10^3$s. As the
photon index is steeper than 2, the emission energy is dominated
by low-energy, $\sim100$~MeV, photons. It could be that the
dominant $>100$~MeV component of $>10^3$s scale and the $>10$~GeV
photon of still earlier arrival time have different origins. On
the other hand, the properties of the $>100$~MeV emission, i.e.,
the spectral index, the light curve slope and the flux level, are
consistent with the external shock model, supporting our
assumption. We also note that the $>100$~MeV flux, locating above
$\nu_c$, is insensitive to the poorly known model parameters, i.e,
$n$ and $\eps_B$ \citep[e.g.,][]{fw01}, giving us confidence on
the external shock synchrotron assumption.

The LAT-detected $>10$~GeV photons are almost impossible to be
produced by synchrotron radiation because the synchrotron cooling
limits the shock acceleration, therefore might need a different
origin \citep{L10,pn10}. Moreover, the bright GRBs show temporal
variabilities in $>100$~MeV light curves, e.g., GRBs 090510
\citep{fermi09c} and 090902B \citep{fermi09b}, also in
contradiction with external forward shock prediction. Within the
framework of the standard internal shock model, the late,
large-radius residual collisions may naturally give an
interpretation \citep{lw08,L10}.

\cite{kb09a,kb09b}, using the external shock model for the
LAT-emission, concluded that the postshock field is
shock-compressed only from the circum burst medium field. This is
because they explained the whole burst, including the burst phase,
by this model, hence need the external shock MeV emission not to
exceed the observed level. Our result suggests that this might be
impossible, i.e, the preshock field is almost certainly amplified
somehow in the external shock interpretation. In this case, the
MeV emission from the external shock in the prompt burst phase
might be naturally suppressed because the MeV-synchrotron emitting
electrons mainly cool by IC scattering the bulk low-energy
photons.

\acknowledgments

The author thanks E. Waxman, X.Y. Wang, and E. Nakar for comments.
This work was partly supported by the National Natural Science
Foundation of China through grant 10843007 and the Foundation for
the Authors of National Excellent Doctoral Dissertations of China.

\begin{appendix}

\section{Analytical derivation}
We present detailed analytical derivation of $Y_\ob$ values here.
Consider slow and fast cooling regimes separately.
\subsection{Slow cooling} We discuss first the dependence of the
relations between break frequencies on the parameter space. We
will use the broken power-law approximations, neglecting the
correction factors $c_i$'s. We assume $p=2$ for simplicity when
$p$ appears in exponents. Let $\nu_m=\nukn_c$ and using eqs.
(\ref{eq:num}), (\ref{eq:nuc}), (\ref{eq:oc}) and
(\ref{eq:slowcoolingYc}), we get the critical $\eps_B$,
\begin{equation}
  \eps_B=\left\{\begin{array}{lll}
    1.1\times10^{-11}\eta^{-2}f_p^2E_{54}^{-1}n_0^{-2}t_3^{-1} &
    (n>1.8\times10^{-4}\eta^{-2}\eps_{e,-1}^{-4}f_p^{-2}E_{54}^{-1}t_3{\rm ~cm}^{-3}; & Y_c>1)\\
    6.1\times10^{-8}\eps_{e,-1}^4f_p^4n_0^{-1}t_3^{-2} &
    (n<1.8\times10^{-4}\eta^{-2}\eps_{e,-1}^{-4}f_p^{-2}E_{54}^{-1}t_3{\rm ~cm}^{-3}; & Y_c<1)
  \end{array}.\right.
\end{equation}
This means
\begin{equation}
  \nukn_c\left\{\begin{array}{l}
    < \\ >
  \end{array}\right\}\nu_m\Leftrightarrow
  \eps_B\left\{\begin{array}{l}
    < \\ >
  \end{array}\right\}\min[1.1\times10^{-11}\eta^{-2}f_p^2E_{54}^{-1}n_0^{-2}t_3^{-1},~6.1\times10^{-8}\eps_{e,-1}^4f_p^4n_0^{-1}t_3^{-2}].
\end{equation}

Let $\nu_c=\nukn_c$ and, again, using eqs. (\ref{eq:num}),
(\ref{eq:nuc}), (\ref{eq:oc}) and (\ref{eq:slowcoolingYc}), we get
another critical $\eps_B$,
\begin{equation}
  \eps_B=\left\{\begin{array}{lll}
    1.0\times10^{-6}\eta^{-6/5}\eps_{e,-1}^{-12/5}f_p^{-6/5}E_{54}^{-1}n_0^{-6/5}t_3^{3/5} &
    (n>1.8\times10^{-4}\eta^{-2}\eps_{e,-1}^{-4}f_p^{-2}E_{54}^{-1}t_3{\rm ~cm}^{-3}; & Y_c>1)\\
    1.8\times10^{-4}E_{54}^{-2/5}n_0^{-3/5} &
    (n<1.8\times10^{-4}\eta^{-2}\eps_{e,-1}^{-4}f_p^{-2}E_{54}^{-1}t_3{\rm ~cm}^{-3}; & Y_c<1)
  \end{array}.\right.
\end{equation}
This means
\begin{equation}\label{eq:ocvsnuc}
  \nukn_c\left\{\begin{array}{l}
    < \\ >
  \end{array}\right\}\nu_c\Leftrightarrow
  \eps_B\left\{\begin{array}{l}
    < \\ >
  \end{array}\right\}\min[1.0\times10^{-6}\eta^{-6/5}\eps_{e,-1}^{-12/5}f_p^{-6/5}E_{54}^{-1}n_0^{-6/5}t_3^{3/5},~1.8\times10^{-4}E_{54}^{-2/5}n_0^{-3/5}].
\end{equation}

Now let us calculate $Y_\ob$ value in the case with the lowest
$\eps_B$ value allowed, $\eps_B=10^{-5}$, and with the other
parameters being $E_{54}=t_3=n_0=\eps_{e,-1}=1$. This case
marginally satisfies the condition for slow cooling regime,
$\eps_B\la\eps_{B,\rm cr}$, and, from eq. (\ref{eq:ocvsnuc}),
satisfies $\nu_c<\nukn_c$ as well. Thus from eq.
(\ref{eq:slowcoolingYc}), we have $
  Y_c(1+Y_c)=10^4\eps_{e,-1}\eps_{B,-5}^{-1}\eta,
$
which gives the solution
\begin{equation}\label{eq:Ycsolution1}
  Y_c=10^2\eta^{1/2}\eps_{e,-1}^{1/2}\eps_{B,-5}^{-1/2}c_3.
\end{equation}
Substitute eq. (\ref{eq:Ycsolution1}) into $\nu_m$ and $\nu_c$
(eqs. \ref{eq:num} and \ref{eq:nuc}), and hence into the
definition $\eta=(\nu_m/\nu_c)^{(p-2)/2}$, then $\eta$ can be
solved out to be
\begin{equation}
  \eta=(0.57E_{54}\eps_{B,-5}\eps_{e,-1}^3f_p^2n_0t_3^{-1})^{(p-2)/(4-p)}.
\end{equation}
For $p=2.2$ and 2.4, $\eta=0.94$ and 0.87, respectively, close to
unity.

With eqs.(\ref{eq:Ycsolution1}) and (\ref{eq:slowcoolingYob}), we
get, for $\eps_B=10^{-5}$ case,
\begin{equation}
  Y_\ob=9.9c_3c_6\eta E_{54}^{1/6}n_0^{1/2}
  \eps_{B,-5}^{-1/3}\eps_{e,-1}^{-2/3}f_p^{-5/3}t_3^{1/2}.
\end{equation}

\subsection{Fast cooling} Again we discuss the parameter space
dependence first. Let $\nukn_\ob=\nu_c$ and using eqs.
(\ref{eq:nuc}), (\ref{eq:oob}) and (\ref{eq:fastcoolYm}), we get
the critical $\eps_B$,
\begin{equation}
  \eps_B=1.6\times10^{-4}\eps_{e,-1}^{-4/3}E_{54}^{-1}n_0^{-4/3}t_3^{1/3},
\end{equation}
where
\begin{equation}
  \nukn_\ob\left\{\begin{array}{l}
    < \\ >
  \end{array}\right\}\nu_c\Leftrightarrow
  \eps_B\left\{\begin{array}{l}
    < \\ >
  \end{array}\right\}1.6\times10^{-4}\eps_{e,-1}^{-4/3}E_{54}^{-1}n_0^{-4/3}t_3^{1/3}.
\end{equation}

Now consider the case of high $\eps_B$ value, $\eps_{B}=10^{-2}$,
with the other parameters being $E_{54}=t_3=n_0=\eps_{e,-1}=1$.
These parameter values obviously satisfies fast cooling condition,
$\eps_B>\eps_{B,\rm cr}$, and that $\nu_c<\nukn_\ob$. From eq.
(\ref{eq:fastcoolingYob}) we write
$Y_\ob=Y_m(\nukn_\ob/\nu_m)^{1/2}c_8$, i.e.,
\begin{equation}
  Y_\ob=0.62c_8E_{54}^{-1/8}\eps_{B,-2}^{-5/8}\eps_{e,-1}^{-1/2}f_p^{-1}t_3^{3/8}.
\end{equation}

\section{Numerical calculation for $\eps_B=10^{-5}$ case}
As the lowest $\eps_B$ case gives the most relaxed lower limit to
the upstream field, we carry more detailed numerical calculation
in this case, in particular we should calculate the whole
synchrotron photon spectrum, considering KN effect, for more
accurate calculation of the IC cooling of electrons. Consider
$\epsilon_e=0.1$, $\epsilon_B=10^{-5}$, $n=1\rm cm^{-3}$,
$E=10^{54}$erg, and $t=10^3$s, and consider the downstream
shocked-plasma frame. Note for simplicity in this section we
neglect primes for the quantities in the downstream frame. As
discussed previously, $\eps_B\sim\eps_{B,\rm cr}$ for taken
parameters, thus $\nu_m\sim\nu_c$, i.e., the system could be in
either slightly fast or slow cooling regime, thus we need to
consider the possibility of both regimes.

Consider first slow cooling regime. We assume $\nu_c<\nukn_c$, as
analyzed in \S\ref{sec:KN}, which can also be checked later (see
below). The electron distribution at $\gamma_m<\gamma<\gamma_c$
still follows the injected form $n_\gamma\propto\gamma^{-p}$. The
distribution beyond $\gamma_c$ is strongly affected by radiative
cooling. By the electron continuity equation, we have
\begin{equation}
  n_\gamma\propto\gamma^{-(p-1)}/\dot{\gamma}.
\end{equation}
The electron energy loss rate might be affected by KN effect,
\begin{equation}
  \dot{\gamma}\propto\gamma^2(1+Y)\propto\gamma^2[u_B+u_{\rm
ph}(<\nukn)].
\end{equation}
So the electron distribution around $\gamma$ depends on the photon
energy distribution around $\nukn\equiv m_ec^2/h\gamma$, and the
electron distribution and synchrotron photon spectrum are coupled.
However, if the photons that affect the electron distribution at a
certain electron energy range are not those radiated by the same
electrons, then the analysis becomes simple. This is the case
here. As $\nu_c<\nukn_c$, the electrons just above $\gamma_c$ have
$u_{\rm ph}(<\nukn)\approx u_{\rm ph,syn}$, independent of
$\gamma$, thus the electrons follow the form
$n_\gamma\propto\gamma^{-(p+1)}$. As $\gamma$ increases to
$\gamma>\hat{\gamma}_c\equiv m_ec^2/h\nu_c$, the electrons mainly
interact with photons at $\nukn<\nu_c$ (where
$u_\nu\propto\nu^{-(p-1)/2}$) hence $u_{\rm
ph}(<\nukn)\propto\gamma^{-(3-p)/2}$, and
$n_\gamma\propto\gamma^{-(3p-1)/2}$. This scaling holds until
$\gamma>\hat{\gamma}_m\equiv m_ec^2/h\nu_m$, beyond which the
electrons mainly interact with photons at $\nukn<\nu_m$ where
$u_\nu\propto\nu^{1/3}$, hence $u_{\rm
ph}(<\nukn)\propto\gamma^{-4/3}$, and
$n_\gamma\propto\gamma^{-(p-1/3)}$. So far we actually assume
$u_{\rm ph}(<\nukn)/u_B>1$, this is true even for electrons with
$\hat{\gamma_m}$ (see below), but not true for electron with high
enough energy, $\gamma>\gamma_B$, where $Y(\gamma_B)=u_{\rm
ph}(<\nukn_B\equiv m_ec^2/h\gamma_B)/u_B=1$. $\nukn_B$ and hence
$\gamma_B$ can be solved since the low energy end of the photon
spectrum is known, $u_\nu\propto\nu^{1/3}$. At $\gamma>\gamma_B$,
the electrons lie in deep KN regime and the energy loss is
dominated by synchrotron radiation, so
$\dot{\gamma}\propto\gamma^2$ and
$n_\gamma\propto\gamma^{-(p+1)}$. Given the electron distribution
above, the synchrotron photon energy density per unit frequency,
for slow cooling regime, can be given as broken power laws,
\begin{equation}\label{eq:apendix:spec1}
  u_\nu\propto\left\{\begin{array}{ll}
     \nu^{1/3}  & \nu<\nu_m\\
     \nu^{-(p-1)/2}  & \nu_m<\nu<\nu_c\\
     \nu^{-p/2} & \nu_c<\nu<\hat{\nu}_c\\
     \nu^{-3(p-1)/4}  &  \hat{\nu}_c<\nu<\hat{\nu}_m\\
     \nu^{-(3p-4)/6}   &  \hat{\nu}_m<\nu<\nu_B\\
     \nu^{-p/2}  &  \nu_B<\nu.
     \end{array}\right.
\end{equation}
Here $\nu_x=\nu_{\rm syn}(\gamma_x)$ and $\hat{\nu}_x=\nu_{\rm
syn}(\hat{\gamma_x})$ with $\nu_{\rm
syn}(\gamma)\equiv0.3\gamma^2eB_d/2\pi m_ec$. Note $\nu_c$ and
$\hat{\nu}_c$ are functions of $\gamma_c$, while $\nu_m$ and
$\hat{\nu}_m$ can be directly calculated for given parameters, and
the discussion of $\nu_B$ is given later.

Next consider the fast cooling case ($\nu_c<\nu_m$). As discussed
in \S\ref{sec:KN}, since $\nukn_m>\nu_m$ the electrons at
$\gamma_c<\gamma<\gamma_m$ follow $n_\gamma\propto\gamma^{-2}$,
and then $n_\gamma\propto\gamma^{-(p+1)}$ just above $\gamma_m$.
At $\hat{\gamma}_m<\gamma<\hat{\gamma}_c$, electrons mainly cool
by photons at $\nu_c<\nukn<\nu_m$ (where $u_\nu\propto\nu^{-1/2}$)
then $u_{\rm ph}(<\nukn)\propto\gamma^{-1/2}$ and
$n_\gamma\propto\gamma^{-(p+1/2)}$. At still larger $\gamma$'s the
electrons cool by photons in $u_\nu\propto\nu^{1/3}$ segment, and
then $n_\gamma\propto\gamma^{-(p-1/3)}$. Again at large enough
$\gamma$ for deep KN regime, synchrotron energy loss is dominant
then $n_\gamma\propto\gamma^{-(p+1)}$ ($\gamma>\gamma_B$). We can
write for fast cooling regime
\begin{equation}\label{eq:appendix:spec2}
  u_\nu\propto\left\{\begin{array}{ll}
     \nu^{1/3}  & \nu<\nu_c\\
     \nu^{-1/2}  & \nu_c<\nu<\nu_m\\
     \nu^{-p/2} & \nu_m<\nu<\hat{\nu}_m\\
     \nu^{-(2p-1)/4}  &  \hat{\nu}_m<\nu<\hat{\nu}_c\\
     \nu^{-(3p-4)/6}   &  \hat{\nu}_c<\nu<\nu_B\\
     \nu^{-p/2}  &  \nu_B<\nu.
     \end{array}\right.
\end{equation}
Note in this case not only $\nu_c$ and $\hat{\nu}_c$ but also
$\nu_B$ is function of $\gamma_c$ since $u_\nu$ peak at $\nu_c$
(see below).

The spectral peak at $\nu_{\max}\equiv\min(\nu_m,\nu_c)$ can be
calculated directly for given parameters,
$
  u_{\nu_{\max}}= n_eP_{\nu_{\max}}t_{\rm ad},
$
where $n_e=4\Gamma n$ is the postshock electron density,
$P_{\nu_{\max}}=\sqrt{3}e^3B_d/m_ec^2$ is the maximum spectral
synchrotron power per unit frequency by one electron, and $t_{\rm
ad}=6R/13\Gamma c$ is the adiabatic cooling time of postshock
plasma. Now discuss $\nu_B$. Assuming $\nukn_B<\min(\nu_m,\nu_c)$
(checked later), $\nukn_B$ is given by
$\int_0^{\nukn_B}u_{\nu_{\max}}(\nu/\nu_{\max})^{1/3}d\nu=u_B$,
and hence $\gamma_B=m_ec^2/h\nukn_B$ and $\nu_B=\nu_{\rm
syn}(\gamma_B)$ are found. $\nu_B$ is a constant in slow cooling
regime but a function of $\gamma_c$ in fast cooling regime.

Thus the only unknown value to determine the whole synchrotron
spectrum is $\gamma_c$, relevant to electrons cool significantly
in a time of $t_{\rm ad}$, i.e.,
$\dot{\gamma}(\gamma_c)=\gamma_c/t_{\rm ad}$, which reads
\begin{equation}\label{eq:appendix:gc}
   \gamma_c(1+Y_c)=\frac{6\pi m_ec}{\sigma_TB_d^2t_{\rm ad}},
   ~~~~~Y_c=\frac1{u_B}\int_0^{\nukn_c}u_\nu d\nu.
\end{equation}

Now we carry numerical calculation to find the root $\gamma_c$ of
eq (\ref{eq:appendix:gc}), with helps of eqs
(\ref{eq:apendix:spec1}) and (\ref{eq:appendix:spec2}). For given
parameters the root turns out to be
$\gamma_c=1.37(1.60)\times10^4$ for $p=2.2(2.4)$. Thus $\gamma_c$
and $\gamma_m(=1.07\times10^4)$ are close, and the system is
marginally in slow cooling regime, consistent with the fact that
$\eps_B\la\eps_{B,\rm cr}$.  With this result, it can be checked
that the assumptions of $\nukn_c>\nu_c$ and $u_{\rm
ph}(<\nu_m)>u_B$ (hence $\hat{\gamma}_m<\gamma_B$) are satisfied.

With $\gamma_c$ known then we have all the break frequencies known
and hence the spectrum $u_\nu$, thus the Compton $Y$ parameter is
given by $Y(\gamma)=(1/u_B)\int_0^{\nukn(\gamma)}u_\nu d\nu$. For
the observed 100-MeV photons, $\gamma_\ob=8.3\times10^7$, thus we
get $Y_\ob=1.37$ (independent of $p$). A more accurate calculation
is taking the KN cross section, $\sigma_{\rm KN}$, and the
relativistic energy transfer in scattering into account,
\begin{equation}\label{eq:Y_KN}
   Y(\gamma)=\frac{3}{8\sigma_T u_B}\int_0^\infty
d\nu u_\nu \int_{-1}^1 d\mu\frac{(1-\mu)^2\sigma_{\rm
  KN}[{\nu\over\nukn(\gamma)}(1-\mu)]}{1+{\nu\over\nukn(\gamma)}(1-\mu)},
\end{equation}
where $\mu=\cos\theta$ and $\theta$ is the angle between the
interacting electron and photon. This gives $Y_\ob=1.18$ for
$p=2.2$ or 2.4. The results are similar to that using
$Y(\gamma)=u_{\rm ph}(<\nukn[\gamma])/u_B$ (eq
\ref{eq:Yapproximation}), indicating that the latter is an
excellent approximation.

\end{appendix}

\end{document}